\documentclass[prd,showpacs,preprintnumbers,amsmath,amssymb,twocolumn,superscriptaddress,notitlepage]{revtex4-2}

\usepackage[utf8]{inputenc}

\usepackage[normalem]{ulem}

\usepackage{graphicx}
\usepackage[usenames,dvipsnames]{color}
\usepackage[
colorlinks=true,       
linkcolor=red,          
citecolor=blue,        
filecolor=magenta,      
urlcolor=cyan  
]{hyperref}

\usepackage{dcolumn}
\usepackage{bm}

\hypersetup{colorlinks,citecolor= blue,linkcolor= blue, urlcolor=blue}

\usepackage{amsmath}
\usepackage{bbm}
\usepackage{amsfonts}
\usepackage{amssymb}
\usepackage{latexsym}
\usepackage{graphicx}
\usepackage[english]{babel}
\usepackage{multirow}
\usepackage{float}
\usepackage{url}
\usepackage{slashed}
\usepackage{xcolor} 
\usepackage{array}
\usepackage{cases}

\newcommand{\be}{\begin{equation}}
\newcommand{\ee}{\end{equation}}
\newcommand{\ba}{\begin{array}}
\newcommand{\ea}{\end{array}}
\newcommand{\bea}{\begin{eqnarray}}
\newcommand{\eea}{\end{eqnarray}}

\usepackage{ulem,fancyvrb}
\usepackage{xcolor}

\begin{document}

\title{
Enhanced Monochromatic Photon Emission from \\ Millicharged Co-Interacting Dark Matter
}

\author{Mingxuan Du}  \email{mingxuandu@pku.edu.cn}
\affiliation{School of Physics and State Key Laboratory of Nuclear Physics and Technology, Peking University, Beijing 100871, China}
\affiliation{Center for High Energy Physics, Peking University, Beijing 100871, China}

\author{Jia Liu} \email{jialiu@pku.edu.cn}
\affiliation{School of Physics and State Key Laboratory of Nuclear Physics and Technology, Peking University, Beijing 100871, China}
\affiliation{Center for High Energy Physics, Peking University, Beijing 100871, China}

\author{Xiao-Ping Wang} \email{hcwangxiaoping@buaa.edu.cn}
\affiliation{School of Physics, Beihang University, Beijing 100083, China}
\affiliation{Beijing Key Laboratory of Advanced Nuclear Materials and Physics, Beihang University,
Beijing 100191, China}

\author{Tianhao Wu} \email{tianhaowu@stu.pku.edu.cn}
\affiliation{School of Physics and State Key Laboratory of Nuclear Physics and Technology, Peking University, Beijing 100871, China}

\begin{abstract}
We study a millicharged co-interacting dark matter scenario, where the primary dark matter constituent is the dark photon $A'$ and the secondary component is the fermion $\chi$. In this model, $\chi$ interacts with $A'$ via a $U(1)'$ interaction while being millicharged with respect to normal photons. Our investigation focuses on the oscillation of $A'$ dark matter into photons within the background of $\chi$ particles, revealing that the $A'-\chi$ scattering rate benefits from a Bose enhancement of the $A'$ final state. As the oscillation production rate is directly linked to the scattering rate, the conversion of $A'$ dark matter into monochromatic photons experiences significant amplification owing to this Bose enhancement, especially when the scattering rate $\Gamma_{\rm sca}$ approaches the dark photon mass $m_{A'}$.
These converted monochromatic photons are detectable through radio telescopes and can induce distortions in the Cosmic Microwave Background (CMB) spectrum. We find that the sensitivity of radio telescopes and the constraints imposed by CMB distortion on the kinetic mixing parameter are notably heightened compared to scenarios without the subdominant millicharged dark matter. 
\end{abstract}

\maketitle

\section{introduction}
\label{introduction}

There exists abundant cosmological and astronomical evidence supporting the existence of dark matter, yet its particle properties remain elusive. Ultralight dark matter (ULDM), encompassing the QCD axion \cite{Peccei:1977ur, Peccei:1977hh, Weinberg:1977ma, Wilczek:1977pj, Preskill:1982cy, Abbott:1982af, Dine:1982ah, Kim:2008hd,Wantz:2009it, Ringwald:2012hr, Kawasaki:2013ae}, axion-like particles (ALPs) \cite{Svrcek:2006yi,Essig:2013lka, Marsh:2015xka, Graham:2015ouw}, and dark photons \cite{Nelson:2011sf, Arias:2012az, Graham:2015rva}, emerges as a well-motivated candidate. It offers resolutions to small-scale conundrums such as the cusp/core problem \cite{Flores:1994gz,Navarro:1996gj,Newman:2012nw}, the missing satellite problem \cite{Klypin:1999uc}, and the too-big-to-fail problem \cite{Boylan-Kolchin:2011qkt,Boylan-Kolchin:2011lmk,Sand:2003bp,Tollerud:2014zha}.
Co-interacting dark matter (CoIDM) \cite{Liu:2019bqw} introduces additional interactions to ULDM, thus tackling small-scale issues within the mass range $m > 10^{-22}$ eV and bypassing Lyman-$\alpha$ constraints that clash with fuzzy DM \cite{Menci:2017nsr, Irsic:2017yje, Armengaud:2017nkf, Kobayashi:2017jcf, Murgia:2018now, Nori:2018pka}.

The original CoIDM model contains a dark photon dark matter (DPDM) $A'$ and a fermion dark matter $\chi$ which interacts with $A'$. We have implemented a mass mixing term between $A'$ and the electromagnetic(EM) photon $A$, resulting in $\chi$ acquiring millicharge under the EM photon. This model is referred to as the millicharged CoIDM model.
In this model, the dark photon $A'$ can oscillate into $A$ during the propagation.
A recent study \cite{Du:2023zlt} highlights that the probability of such conversions, i.e. the dark photon production rate via photon oscillation in medium, can be enhanced via the frequent scattering between photon and medium.
Vice versa, the $A'$ conversion to $A$ will also benefit from the scattering between $A'$ and the medium $\chi$. This scattering exhibits the Bose enhancement effect, which greatly enhances the probability of $A'$ conversion to photons.
Therefore, the experiments searching for the monochromatic photons from the above conversion can set stringent limits on the kinetic mixing parameter of $A'$.

In Sec.~\ref{sec: model}, we construct a millicharged CoIDM where the dark photon and EM photon of standard model (SM) are coupled through mass mixing, and a vector interaction between $\chi$ and dark photon is incorporated. 
In Sec.~\ref{sec: dark matter setup}, we discuss the chosen benchmark point parameters.
In Sec.~\ref{Bose enhancement}, we examine the frequent scattering between $A'$ and $\chi$ particles, emphasizing the presence of the Bose enhancement effect in our millicharged CoIDM model.
Sec.~\ref{oscillation probability} briefly reviews the calculation of dark photon oscillations to photons.
In Sec.~\ref{photon production rate} we calculate the photon production rate using the density matrix method and compare it with the previous oscillation probability method. 
We demonstrate that the photon production rate is proportional to the $A'$-$\chi$ scattering rate, reaching its maximum when the scattering rate approaches the dark photon mass.
In Sec.~\ref{search photon from DPDM by radio telescope}, we show how to explore the sensitivity of radio telescopes to the millicharged CoIDM. 
In Sec.~\ref{result and discussion}, we highlight that our model's sensitivity from radio telescopes such as SKA and LOFAR, surpasses existing experimental limits by $6\sim 12$ orders of magnitude.
We also investigate the constraints imposed by Cosmic Microwave Background (CMB) spectral distortion in our millicharged CoIDM model.
Finally Sec.~\ref{conclusion} summarizes the key findings and potential future research directions.

\section{model}
\label{sec: model}

We consider a two-component dark matter model, with the dominant component being the light dark photon $A'$ and the subdominant component being the Dirac fermion $\chi$ charged under the $U(1)_d$ gauge.
The total dark matter abundance is given by the sum of the fractions of relic abundance for the two components, $f_{A'} + f_{\chi} = 1$. 
The corresponding low-energy effective Lagrangian is given by
\be
\mathcal{L} = \mathcal{L}_{\rm SM}
-\frac{1}{4} {F'}_{\mu \nu}^2
+\frac{m_{A'}^2}{2} (A'_\mu - \epsilon A_\mu)^2
+ g_d A'_\mu \bar{\chi} \gamma^\mu \chi.
\label{eq: MM lag}
\ee
Here, ${F'}_{\mu\nu}$ represents the $U(1)_d$ field tensor, while $A$ represents the EM photon. 
The parameter $m_{A'}$ corresponds to the mass of the dark photon, $\epsilon$ denotes the dimensionless mixing parameter, and $g_d$ represents the coupling between $A'$ and $\chi$. 
The mass-mixing term can be generated within the Stueckelberg mechanism~\cite{Feldman:2007wj}, which can lead to millicharged subdominant dark matter component $\chi$.

It should be noted that the fields $(A',\, A)$ in the Lagrangian Eq.~\eqref{eq: MM lag} are the interaction eigenstates, with $A$ being sterile to the dark current of $\chi$. 
The mass eigenstates $(\Tilde{A'},\, \Tilde{A})$, where the photon $\Tilde{A}$ is massless, can be obtained through the following field transformations
\be
A'_\mu \to \frac{\Tilde{A'}_\mu}{\sqrt{1+\epsilon^2}}  
+ \frac{\epsilon \Tilde{A}_\mu}{\sqrt{1+\epsilon^2}} ,\,
A_\mu \to \frac{\Tilde{A}_\mu}{\sqrt{1+\epsilon^2}}  - \frac{\epsilon \Tilde{A'}_\mu}{\sqrt{1+\epsilon^2}}.
\label{eq:massTointer}
\ee
The Lagrangian in the mass basis becomes
\bea
\mathcal{L} &\supset& -\frac{1}{4} \Tilde{F}_{\mu \nu}^2
-\frac{1}{4} \Tilde{F'}_{\mu \nu}^2
+\frac{m_{A'}^2}{2} (1+\epsilon^2) \Tilde{A'}_\mu^2 \nonumber \\
&&+ \frac{g_d}{\sqrt{1+\epsilon^2}} (\Tilde{A}'_\mu + \epsilon \Tilde{A}_\mu )\bar{\chi} \gamma^\mu \chi,
\label{eq: MM lag mass basis}
\eea
where $\Tilde{F}_{\mu \nu}$ ($\Tilde{F'}_{\mu \nu}$) is the field strength tensor of the (dark) photon field in mass basis.
Consequently, the hidden particle $\chi$ is millicharged with an electric charge give by:
\be
Q_\chi = \frac{\epsilon g_d}{e \sqrt{1+\epsilon^2}},
\label{eq: mQ}
\ee
where $e$ represents the absolute magnitude of the charge carried by the electron.

In our analysis, we do not consider the case of the hidden fermion $\chi$ in the dark photon portal, which couples with the EM photon via kinetic mixing. 
Because in this particular case, $\chi$ is not a millicharged particle, as discussed in \cite{Feldman:2007wj, Du:2023hsv, Feng:2023ubl}.
As a result, only the negligible density of SM free charged particles in the universe contributes very small off-diagonal terms $\epsilon m_\gamma^2$ to the mass matrix of photons and dark photons, where $m_\gamma$ is the effective mass of photon in galaxy.
Consequently, the probability of of dark photons converting into photons in space is significantly suppressed by a factor of $m_\gamma^4$.

\section{dark matter setup}
\label{sec: dark matter setup}

We adopt the assumption outlined in Ref.~\cite{Liu:2019bqw}, where the relic abundance of the dark photon $A'$ is attained through nonthermal processes in the early Universe, while the dark fermion $\chi$ is generated via freeze-in mechanism with SM particles. Consequently, we consider the $ee \leftrightarrow \chi \chi$ process to estimate the number density of $\chi$. The relevant cross section is $\left<\sigma v\right> \sim \alpha g_d^2 \epsilon^2 / T^2$ for $T \gg m_{\chi}$. The Boltzmann equation governing the number density $n_{\chi}$ is as follows:
\be
\frac{dn_{\chi}}{dt} + 3Hn_{\chi} = \left<\sigma v\right> n_e^2,
\label{eq: Boltzmann eq}
\ee
where $H \sim T^2/M_{\rm{pl}}$ is the Hubble parameter during the radiation-dominated epoch with Planck mass $M_{\rm{pl}}$, and $n_e \sim T^3$ is the number density of electrons in the thermal bath. By substituting the cross section into Eq.~\eqref{eq: Boltzmann eq}, using the relation $dt = -dT/(HT)$, and integrating up to $T={\rm Max}[m_{\chi},m_{e}]$, we obtain the freeze-in yield for $\chi$
\be
\frac{n_{\chi}}{T^3} \bigg\rvert _{T={\rm Max}[m_{\chi},m_{e}]} \sim \; \alpha  g_d^2  \epsilon^2 \; \frac{M_{\rm{pl}}}{{\rm Max}[m_{\chi},m_{e}]}.
\label{appeq: FI yield}
\ee
Observations indicate the density ratio of dark matter to baryonic matter around 5 \cite{Planck:2018vyg}. 
This leads to $n_{\chi} m_{\chi}/(n_{b} m_{b}) \sim n_{\chi} m_{\chi}/(m_{b}\eta_{b}T^3) \sim 5 f_{\chi}$, where $n_b$ is the baryon number density, $m_b$ is the average mass of baryons, and $\eta_{b} \simeq 6 \times 10^{-10}$ is the baryon-to-photon ratio \cite{Planck:2018vyg}. 
Combining all factors, we obtain 
\be
f_{\chi} \sim 10^{-5} \; g_d^2 \; \left(\frac{\epsilon}{10^{-14}}\right)^2 \; 
\frac{m_\chi}{{\rm Max}[m_{\chi},m_{e}]}.
\label{eq: fchi FI}
\ee
In our setup, we consider the mass of dark photon around $m_{A'} \sim 10^{-6}$ eV, which falls within the detection range of radio telescopes such as SKA \cite{dewdney2013ska1} and LOFAR \cite{Nijboer:2013dxa}. The experimental limit on $\epsilon$ for such mass range is currently around $10^{-14}$. To ensure a significant number density $n_\chi$ while respecting the Pauli exclusion principle, we focus on the mass range of keV $\lesssim m_\chi \lesssim m_e \sim$ MeV.
Furthermore, to avoid the constraints on self-interaction of dark matter, such as the Bullet Cluster, we set the interaction strength between $A'$ and $\chi$ to be $g_d = 0.1 (0.001)$ for MeV(keV) $\chi$ and  (see a more detailed discussion in Sec.~\ref{result and discussion}).
Finally, we choose two benchmark points based on Eq.~\eqref{eq: fchi FI}
\begin{align}
(m_{\chi},f_{\chi},g_d) &= (1\;\rm{keV},2\times10^{-14},0.001), \nonumber \\
(m_{\chi},f_{\chi},g_d) &= (1\;\rm{MeV},1\times10^{-7},0.1).
\end{align}

\section{Co-interacting DM and Bose enhancement.}
\label{Bose enhancement}

Within an environment full of $A'$ and $\chi$ particles, the primary mechanism driving decoherence in the $A'-A$ oscillating system is the Compton-like scattering $A' (p) + \chi (k) \to A' (p') + \chi(k')$, where the momenta of the involved particles are denoted within parentheses. The scattering rate for a $A'$ particle can be calculated as
\begin{eqnarray}
\Gamma_{\rm sca} (p) &\simeq& n_\chi v_{\rm rel} \int d^3 p' \frac{d \sigma_{\rm vac}}{d^3 p'} \left(1+ f_{A'}(p') \right) \nonumber \\
&\approx& n_\chi v_{\rm rel} \sigma_{\rm vac} \left(1+ f_{A'}(p)\right),
\label{eq: Gamma sca}
\end{eqnarray} 
where $n_\chi$ represents the number density of $\chi$, and the relative velocity between $A'$ and $\chi$ is approximately $v_{\rm rel} \sim v_0$, where $v_0 \sim 220~{\rm km/s}$ represents the typical velocity of $A'$ dark matter, while the velocity of $\chi$ is negligibly small as discussed below.
The term $\sigma_{\rm vac}$ corresponds to the Compton scattering cross-section of $A'-\chi$.
More importantly, the factor $1+ f_{A'}(p')$ clearly shows the Bose enhancement from the final states of $A'(p')$ because the $p'$ naturally falls into the momentum range of $A'$ dark matter.

Besides the scattering process, there might be the bremsstrahlung 
$\chi \chi \to \chi \chi A'$ and the inverse bremsstrahlung process $\chi \chi A' \to \chi \chi $ (absorption), which could be efficient to change the number of $A'$ in the late universe \cite{Chang_2019}. 
However, within our relevant parameter space ($g_d =$ 0.1 or 0.001), the ratio between the bremsstrahlung process and the scattering process is quite small, around $10^{-9} \sim 10^{-3}$. In addition, the absorption process of $A'$ does not experience the final state Bose enhancement.
Therefore, we will neglect the bremsstrahlung process and focus on the scattering process between $A'$ and $\chi$ in our further analysis.

Lastly, besides the scattering process, there can be the inverse of the scattering process $A' (p') + \chi(k') \to A' (p) + \chi (k) $. Taking into account this inverse scattering, the effective rate of change for $A'$ in the momentum $p$ state is approximately $\Gamma_{\rm eff} \approx \Gamma_{\rm sca}(p) \times m_{A'}/m_\chi$~\cite{Liu:2019bqw}. However, since we are considering oscillation production of $A$, the $A'$ particles in the final states of both scattering and inverse scattering are new particles with refreshed phases. They will independently contribute to the oscillation production $A' \to A$.  Therefore, in this study, we will use the scattering rate in Eq.~\eqref{eq: Gamma sca}, rather than the effective scattering rate in Ref.~\cite{Liu:2019bqw}.

Coming back to the scattering rate calculation in Eq.~\eqref{eq: Gamma sca}, we assume that dark matter $\chi$ is almost static $ v_\chi = 0$, and the $A'$ state density function $ f_{A'}$ depends only on the magnitude of momentum  $p$. With these assumptions, the integration can be simplified to the second line.
For the first assumption, it can be naturally satisfied in a co-moving volume of $A'$ and $\chi$. Since $A'$ and $\chi$ has a large scattering rate benefited from the Bose enhancement, $A'$ and $\chi$ particles can be in kinetic equilibrium. Given the mass hierarchy $m_{A'} \ll m_\chi$, with both $A'$ and $\chi$ are non-relativistic at the beginning, the kinetic equilibrium velocity of $\chi$ will be negligible comparing to $A'$. However, the final momentum of dark photons will have negligible change in its magnitude due to the total energy conservation after scattering, because $\chi$ has minor mass fraction comparing with $A'$.

It is interesting to compare the the Bose enhancement factor in the $A'$ scattering with the previous neutrino oscillation calculation in the dense medium~\cite{Venumadhav:2015pla}. 
In the case of the sterile neutrino production, the rate at which active neutrinos $\nu_\alpha$ converts to sterile neutrinos $\nu_s$ through oscillation, has the form of $f_ {\nu_\alpha} \Gamma_{\nu_\alpha} P_{\rm osc} (\nu_\alpha \leftrightarrow \nu_s)/2$. The neutrino scattering rate in the dense medium is $\Gamma_{\nu_\alpha} $, which incorporates the the Fermi blocking factor $1-f_{\nu}$~\cite{Venumadhav:2015pla}. Therefore, both our calculation and the sterile neutrino calculation includes the Bose enhancement or Fermi blocking factor in the scattering process for the oscillation production calculation.

Finally, we come back to the density function of $A'$ DM, which has a large scattering rate with the sub-dominant DM component $\chi$. The DPDM is in a kinetic equilibrium with the other components and its number density $n_{A'}$ is given by \cite{Egana-Ugrinovic:2021gnu}: 
\begin{equation}
n_{A'}  \equiv \int \frac{d^3 p}{(2\pi)^3} f_{A'} (p).
\label{eq: kinetic eqb}
\end{equation}
The density function $f_{A'} (p)$ in momentum space follows the Maxwell-Boltzmann distribution of dark matter,
\begin{equation}
f_{A'} (v) \simeq n_{A'} \left(\frac{2 \sqrt{\pi}}{m_{A'} v_0}\right)^{3}e^{- v^2/v_0^2} \equiv \langle f_{A'} \rangle e^{- v^2/v_0^2},
\label{eq:f-p-velocity}
\end{equation}
where we have non-relativistic momentum $p = m_{A'} v$ and the typical velocity of dark matter $v_0$. We can evaluate the averaged density function below
\begin{equation}
\langle f_{A'} \rangle \approx 10^{29} \left(\frac{\rho_{\rm{DM}}}{0.4 \rm{GeV/cm^3}}\right) \left(\frac{v_{0}}{220 \rm{km/s}}\right)^{-3} \left(\frac{m_{A'}}{1 \rm{\mu eV}}\right)^{-4}.
\end{equation}
It shows that for each momentum state, there are a huge number of $A'$ in the same state. Therefore, we can expect the Bose enhancement of the scattering state is at the order of $\langle f_{A'} \rangle$.

\section{Oscillation probability}
\label{oscillation probability}

In this section, we calculate the probability of an initially generated dark photon $A'$ propagating a distance $\ell$ in the universe and eventually converting into a photon in the interaction eigenstates. This conversion can be translated into mass eigenstates using Eq.~\eqref{eq:massTointer}. The Hamiltonian governing $A'$ and $A$ of the transverse mode is given by
\bea
H_{\rm osc}=\frac{1}{2 E_{A'}}\left(\begin{array}{cc}
m_{A'}^2 + \delta m_{A'}^2 & -\epsilon m_{A'}^2 \\
-\epsilon m_{A'}^2 &  \epsilon^2 m_{A'}^2+m_{\gamma}^2
\end{array}\right),
\label{eq: MM int Ham}
\eea
where $E_{A'}$ represents the energy carried by $A'$, $m_{\gamma}=\sqrt{4 \pi \alpha n_e / m_e}$ is the effective masses of photon, and $\delta m_{A'}=\sqrt{g_d^2 n_\chi / m_\chi}$ is the effective mass of dark photon by the medium $\chi$, respectively. $\alpha$ is the fine structure constant, and $n_e\sim 0.015 \, \text{cm}^{-3}$ ($n_\chi$) represents the number density of electrons ($\chi$) in galaxy \cite{Ocker:2020tnt}.
The typical value of the effective masses in the galactic DM halo are approximately
\bea
m_{\gamma} \simeq \; &5& \times 10^{-12} \; {\rm{eV}} \times \left(\frac{n_e}{0.015 \, \rm{cm^{-3}}}\right)^{1/2} ,  \\
\delta m_{A'} \simeq \; &5& \times 10^{-20} \; {\rm{eV}} 
\times \left(\frac{g_d}{0.1}\right) 
\times \left(\frac{f_\chi}{10^{-7}}\right)^{1/2} 
\times \left(\frac{{\rm MeV}}{ m_\chi }\right). \nonumber
\label{eq: meff}
\eea
We can find that the medium-induced effective masses on the photon and dark photon are both negligible compared to $m_{A'}$ for our benchmark point.

When dark photons propagate through a medium, the effect of the medium on the oscillation Hamiltonian Eq.~\eqref{eq: MM int Ham} can be accounted for by the substitution  $m_{A'}^2 \to m_{A'}^2 + i E_{A'} \Gamma_{A'}$, where $\Gamma_{A'}$ represents the damping rate for dark photons in the medium. 
If a dark photon travels a distance $\ell$ within the medium, the transition probability of the interaction eigenstate from $A'$ to $A$ can be calculated from the modified Hamiltonian~\cite{Redondo:2015iea}:
\begin{align}
P_{\rm osc} & \left( t = \ell/v\right) = \epsilon^2 m_{A'}^4  \nonumber \\
& \times \frac{1+e^{-\Gamma_{A'} \ell/v}-2 \cos \frac{m_{A'}^2 \ell}{2 E_{A'} v} e^{-\Gamma_{A'} \ell/(2v)}}{m_{A'}^4+(E_{A'} \Gamma_{A'})^2}.
\label{eq: osc prob}
\end{align}
In the long-distance limit $\ell \to \infty$, it recovers the well-known result \cite{Redondo:2008aa, An:2013yfc, Redondo:2013lna, Redondo:2015iea}
\begin{equation}
P_{\rm osc}^{\infty} = \epsilon^2 \frac{ m_{A'}^4 }{m_{A'}^4+(E_{A'} \Gamma_{A'})^2},
\label{eq:osc-prob-infty}
\end{equation}

In general, the damping process encompasses all processes where the momentum of dark photons is altered by the medium, including absorption and scattering. 
In our model, we focus solely on the Compton-like $A'-\chi$ scattering process as the primary damping process since absorption rate is significantly lower than the scattering rate as discussed in Sec.~\ref{Bose enhancement}.


\section{photon production rate}
\label{photon production rate} 

To calculate the production rate of $A$ contributed by the oscillation from dark photon $A'$ to $A$ in the interaction basis, we start by considering a single dark photon with momentum $p$ generated from $A'-\chi$ scattering. The evolution can be described using the density matrix method as follows \cite{Sigl:1993ctk}
\begin{equation}
\dot{\rho}_p =-\mathrm{i}[H_{\rm osc}, \rho_p] 
- \sum_{p' \neq p} \frac{1}{2}\left\{G_{p \to p'},  \rho_p \right\},
\label{eq: dot-rho-p}
\end{equation}
where $H_{\rm osc}$, represents the oscillation Hamiltonian, while the second term represents the damping effect of the dark photon. 
Considering the medium is filled with $A'$ and $\chi$ particles, and the dominant damping process is Compton-like scattering between $A'-\chi$, the damping matrix in Eq.~\eqref{eq: dot-rho-p} can be written as
\begin{equation}
\sum_{p' \neq p} G_{p\to p'} \equiv {\rm{Diag}}\left[ \Gamma_{\rm sca},\,0 \right].
\end{equation}
Here $\Gamma_{\rm sca}$, given by Eq.~\eqref{eq: Gamma sca}, represents the scattering rate of a single dark photon in an environment full of $A'$ and $\chi$ particles.

To solve Eq.~\eqref{eq: dot-rho-p}, we parameterize the single particle density matrix as
\begin{eqnarray}
\rho_p =\left(\begin{array}{cc}
1 & 0 \\
0 & 0
\end{array}\right)+
\left(\begin{array}{cc}
h_{A'} & g \\
g^* & h_A
\end{array}\right),
\label{eq: para}
\end{eqnarray}
where the first matrix on the right is the initial state of the density matrix and the second matrix describes the deviation from the initial state.
Note that the time scale of Eq.~\eqref{eq: dot-rho-p} is the mean free time of dark photons in the medium $1/\Gamma_{\rm sca}$. 
Consequently, the leading order of $\rho_{11}$ is 1, and it is natural to have 
Substituting Eq.~\eqref{eq: para} into Eq.~\eqref{eq: dot-rho-p}, we obtain
\begin{eqnarray}
\dot{h}_A &=& 2 H_{\rm osc, 12} {\rm Im} g \nonumber \\
\dot{h}_{A'} &\simeq&  - 2 H_{\rm osc, 12} {\rm Im} g -\Gamma_{\rm sca}  \\
\dot{g} &\simeq& 
-\left( \frac{\Gamma_{\rm sca}}{2} 
+ i (H_{\rm osc, 11} - H_{\rm osc, 22}) \right) g 
+ i  H_{\rm osc, 12}. \nonumber 
\label{eq: diff}
\end{eqnarray}
With initial condition $g (t =0) = 0$ as in Ref.~\cite{Sigl:1993ctk, Redondo:2013lna}, one integrate $\dot{g}$ over time
\begin{equation}
g(t) =  \epsilon m_{A'}^2 \frac{1 - e^{-\frac{i m_{A'}^2 + \Gamma_{\rm sca} E_{A'}}{2 E_{A'}} t}}{ m_{A'}^2 - i  E_{A'} \Gamma_{\rm sca}} .
\end{equation}
For large $t$,
the production rate of normal photon $A$ from single $A'$ by the oscillation is given by
\begin{equation}
\dot{\rho}_{p, 22} = \dot{h}_A \approx P_{\rm osc}^{\infty} \Gamma_{\rm sca}.
\end{equation}
Therefore, the production rate of $A$ with momentum $p$ is given by
\begin{equation}
\dot{f}_{A} = f_{A'}(p) P_{\rm osc}^{\infty} \Gamma_{\rm sca},
\label{eq: dot fA DenMat}
\end{equation}
where $f_{A'}(p)$ is the density of dark photon with momentum $p$.

We can understand above result from another microscopic perspective, the photon production rate $\dot{f}_{A}$ can be estimated as the sum of the rates at which dark photons oscillate into photons before the dark photon is scattered off by $\chi$ particles:
\begin{equation}
\dot{f}_A(p) = f_{A'}(p) \frac{P_{\rm osc} (t_{\rm free})}{t_{\rm free}} \approx f_{A'}(p) P_{\rm osc}^{\infty} \Gamma_{\rm sca},
\label{eq: dot fA micro}
\end{equation}
where $t_{\rm free} \approx 1/\Gamma_{\rm sca}$ represent the mean free time for $A'$ scattering to $\chi$.

At the end of this section, we discuss the behavior of $\dot{f}_A$ depending on changes in $\Gamma_{\rm sca}$.
For simplicity, we drop the exponential terms in the numerator of Eq.~\eqref{eq: osc prob}, which leads to order one correction.
The production rate $\dot{f}_{A}$ can be approximately given by: 
\begin{equation}
\dot{f}_A \approx f_{A'}(p) \frac{\epsilon^2 m_{A'}^4}{m_{A'}^4 + m_{A'}^2 \Gamma_{\rm sca}^2} \Gamma_{\rm sca}.
\end{equation}
When $\Gamma_{\rm sca} = 0$, there is no scattering between dark photons and $\chi$, thus there is no generation of the interaction eigenstate $A'$ through scatterings. 
Consequently, in a homogeneous universe, all dark photons are currently in mass eigenstates, preventing their conversion into photons.

In the limit of $\Gamma_{\rm sca} \ll m_{A'}$, the oscillation production rate of photons can be simplified to 
\begin{equation}
    \dot{f}_A \to f_{A'}(p) \times \epsilon^2 \Gamma_{\rm sca},
    \label{eq:photonproduction-simplified}
\end{equation}
where the second term $\epsilon^2 \Gamma_{\rm sca}$ presents the usual expectation from the conversion process $A' \chi \to A \chi$. For example, the conversion process should have a smaller rate by the factor $\epsilon^2$ comparing with the scattering process $A' \chi \to A' \chi$, due to the suppressed millicharge coupling between $\chi$-$A$. 
More importantly, the scattering term $\Gamma_{\rm sca}$ in Eq.~\eqref{eq:photonproduction-simplified}
contains the Bose enhancement in the scattering process, which enlarges the 
oscillation production of photons through large scattering. 
This results coincides with our previous work~\cite{Du:2023zlt}, that the photon production is enhanced by the scattering process when the scattering rate is larger than the absorption rate. 
\textit{The difference in this scenario is that the scattering rate contains extra Bose enhancement from the $A'$ final state.} Therefore, we have shown that the result from density matrix calculation is in agreement with our previous study~\cite{Du:2023zlt}.

In another limit $\Gamma_{\rm sca} \gg m_{A'}$, we have the photon production $\dot{f}_A \to \epsilon
^2 f_{A'}(p)/\Gamma_{\rm sca}$, which is suppressed by too large scattering rate.  
This means that extremely frequent collisions disrupt the coherence required for the oscillation evolution, and the Quantum Zeno effect appears which suppress the photon production rate~\cite{Arguelles:2016uwb}. In Ref.~\cite{Arguelles:2016uwb}, the similar scenario has been discussed in sterile neutrino - neutrino oscillation case, where the Quantum Zeno effect happens when $E_\nu \Gamma_{\rm sca} \gg \left(m_{\rm sterile}^2 - m_{\nu}^2 \right)$. In our study, the non-relativistic dark matter $A'$ has an energy close to its mass, thus leading to the simplified criterion $\Gamma_{\rm sca} \gg m_{A'}$.
Lastly, it is worth mentioning when $\Gamma_{\rm sca} \approx m_{A'}$, the photon production rate reaches its maximum, and experiments have the best sensitivity to these parameter spaces.

\section{photon flux density by radio telescope}
\label{search photon from DPDM by radio telescope}

We can estimate the photon luminosity along the light of sight (l.o.s.) resulting from DPDM oscillation in our galaxy. Including the conversion probability and Bose enhancement, similar to solar dark photon emission~\cite{Redondo_2015}, the photon luminosity can be written as
\begin{equation}
\mathcal{L}_{\gamma} = 
\int \frac{d r}{4 \pi}
\int \frac{d^3 p}{(2\pi)^3} f_{A'}(p) P_{\rm osc} \Gamma_{\rm sca} E_{A'} 
\label{eq: gamma lum}
\end{equation}
where $r$ is the distance between DPDM to the detector,
the density distribution of DPDM is assumed to follow the NFW profile \cite{Navarro:1995iw},
and $E_{A'} \simeq m_{A'}$ is the energy of the non-relativistic DPDM.

The signal flux density in a radio telescope is calculated as
\be
S= \frac{\mathcal{L}_\gamma}{B},
\label{eq:S-telescope}
\ee
where $B = {\rm max}(B_{\rm res}, v^2 m_{A'} / (4\pi))$ represents the frequency bandwidth. 
Here, $B_{\rm res}$ corresponds to the telescope spectral resolution. 
In the parameter space we are interested, it approximately satisfies $B \simeq B_{\rm res}$.

For a radio telescope, the minimum detectable flux density is given by \cite{An:2020jmf}
\be
S_{\min }=\frac{2 T_{\rm sys}}{A_{\rm eff} \eta_s \sqrt{n_{\rm{pol}} B t_{\rm{obs}}}},
\label{eq:S-min}
\ee
where $T_{\rm sys}$ represents the antenna system temperature, $A_{\rm eff}$ is the effective area of the antenna array, $\eta_s = 0.9$ (1) for SKA \cite{dewdney2013ska1} (LOFAR \cite{Nijboer:2013dxa}), $\eta_{\rm pol} = 2$ denotes the number of polarizations, and $t_{\rm obs} = 1$ hour represents the observation time in the subsequent analysis.
Tab.~\ref{tab: radio telescope}, provides details on the frequency range, telescope spectral resolution, and minimum detectable flux density \cite{An:2020jmf,Braun:2019gdo} for the SKA and LOFAR radio telescopes.

\begin{table}[htbp]
\begin{tabular}{lccc}
\hline \hline Name & $f(\rm{MHz})$ & $B_{\rm res} ({\rm kHz})$ & $S_{\text {min}}(\rm{Jy})$ \\
\hline LOFAR & $[10,80]$ & 195 & $ 1.1$   \\
LOFAR & $[120,240]$ & 195 & $ 8.5\times 10^{-2}$  \\
SKA1-low & $[50,350]$ & 1 & $ 3.5\times 10^{-3}$  \\
SKA1-mid B1 & $[350,1050]$ & 3.9 & $ 6.6\times 10^{-4}$  \\
SKA1-mid B2 & $[950,1760]$ & 3.9 & $ 3.8\times 10^{-4}$   \\
SKA1-mid B5a & $[4600,8500]$ & 3.9 & $ 5.2\times 10^{-4}$  \\
SKA1-mid B5b & $[8300,15300]$ & 3.9 & $ 6.7\times 10^{-4}$   \\
\hline \hline
\end{tabular}
\caption{frequency range, telescope spectral resolution, and the minimum detectable flux density for SKA1 and LOFAR.
The values are taken from Ref.~\cite{An:2020jmf,Braun:2019gdo}}
\label{tab: radio telescope}
\end{table}

\section{result and discussion}
\label{result and discussion}

\begin{figure}[htb]
\begin{centering} 
\includegraphics[width=0.98 \columnwidth]{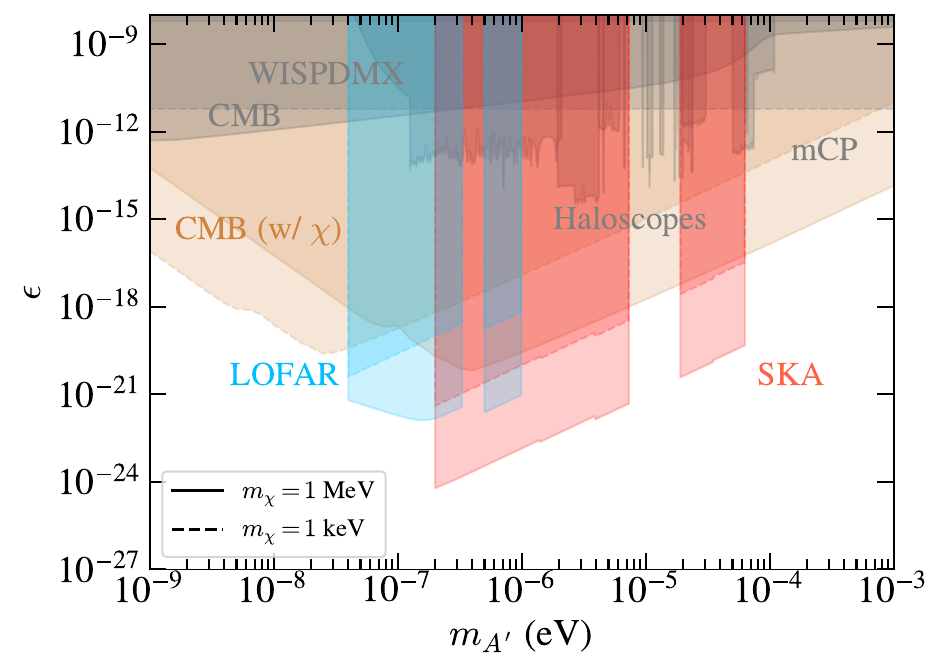}
\caption{LOFAR (blue) and SKA1 (red) sensitivities on the millicharged Co-interacting dark matter model.
The solid and dashed lines indicate the benchmark points of $(m_{\chi},f_{\chi},g_d) = (1\;\rm{MeV},1\times10^{-7},0.1)$ and $(m_{\chi},f_{\chi},g_d) = (1\;\rm{keV},2\times10^{-14},0.001)$.
The light brown regions are the constraints of CMB spectral distortion to our co-interacting DPDM.
The current constraints denoted by gray shaded regions are obtained from the existing haloscope dark photon and axion searches\cite{DePanfilis:1987dk, Wuensch:1989sa, Hagmann:1990tj, ADMX:2001dbg,  ADMX:2009iij}, WISPDMX DPDM searches\cite{Nguyen:2019xuh}, the CMB bounds for DPDM\cite{Arias:2012az,McDermott:2019lch}, and millicharged particle constraints \cite{Davidson:2000hf, Vogel:2013raa}.
}
\label{fig: result}
\end{centering}
\end{figure}

We estimate the sensitivity by applying the condition $S > S_{\rm min}$, based on Eq.(\ref{eq:S-telescope}) and Eq.(\ref{eq:S-min}). We consider two benchmark points: $(m_{\chi},f_{\chi},g_d) = (1\;\rm{MeV},1\times10^{-7},0.1)$ (dashed line) and $(m_{\chi} ,f_{\chi},g_d) = (1\;\rm{keV},2\times10^{-14},0.001)$ (solid line), as discussed in Sec.~\ref{sec: dark matter setup}.
Fig.~\ref{fig: result} illustrates the sensitivity of the SKA (red) and LOFAR (blue) radio telescopes to our millicharged CoIDM model for an observation duration of one hour,  pointing to the direction perpendicular to the Galactic plane.
The gray shaded region in Fig.~\ref{fig: result} represents the current limit for DPDM and millicharged particles.
Our results demonstrate that LOFAR sensitivity surpasses the existing limitations by 6 to 9 orders of magnitude in the mass range of $m_{A'} = 4 \times 10^{-8}$ to $10^{-6}$ eV. 
Additionally, SKA exhibits a remarkable improvement, with a sensitivity that is 6 to 12 orders of magnitude stronger than the existing limit in the mass range of $m_{A'} = 2 \times 10^{-7}$ to $7 \times 10^{-6}$ eV.

The shapes of radio telescope sensitivity can be explained as follows. 
Given the telescope is oriented 
perpendicular to the Galactic plane, 
the photon flux along l.o.s. is dominated by the local dark matter due to the significantly reduced dark matter density at long distances. 
Therefore, we can focus on the local dark matter. 
As the discussion in the end of Section.~\ref{photon production rate}, the photon production rate reaches its maximum when $\Gamma_{\rm sca}\sim m_{A'}$. Thus we can get the highest sensitivity after applying this condition to Eq.~\eqref{eq: gamma lum}. Any deviations of $\Gamma_{\rm sca}$ from $m_{A'}$ results in a weakening of the sensitivity.
Consequently, for the two benchmark points of $(m_{\chi},f_{\chi},g_d) = (1\;\rm{MeV}, 1\times10^{-7},0.1)$ and $(m_{\chi},f_{\chi},g_d) = (1\;\rm{keV}, 2\times10^{-14},0.001)$, we expect the strongest sensitivity to occur at $m_{A'}$ values around $1.7 \times 10^{-7}$ eV and $1.2 \times 10^{-8}$ eV, respectively. 
These values align well with the the turning point of the solid and dashed (out of frequency reach) lines in radio telescope sensitivities from full l.o.s. calculation.

We also explore the constraints imposed by CMB distortion in the millicharged CoIDM model.
In this model, the large scattering rate of the dark photon enhances the process of DPDM oscillating into photons. 
leading to a stronger distortion in the CMB.
The COBE/FIRAS \cite{Fixsen:1996nj} places constraints on this spectral distortion,  quantified by $y$- and $\mu$-type parameters \cite{zeldovich1969interaction, sunyaev1970interaction, McDermott:2019lch},
\begin{eqnarray}
y &\simeq& \frac{1}{4}
\int_{T_0}^{5\times 10^4 T_0} \frac{d T}{H T} \frac{J_y }{\rho_{\gamma}} 
\int \frac{d^3 p}{(2\pi)^3} f_{p} P_{\rm osc}  \Gamma_{\rm sca} m_{A'},  \\
\mu &\simeq& 1.401
\int_{5 \times 10^4 T_0}^{2 \times 10^6 T_0} \frac{d T}{H T} \frac{J_{bb} J_\mu}{\rho_{\gamma}} 
\int \frac{d^3 p}{(2\pi)^3} f_{p} P_{\rm osc}  \Gamma_{\rm sca} m_{A'}, \nonumber
\end{eqnarray}
where $T_0$ is today CMB temperature $2.7$ K \cite{Fixsen:2009ug},
$H$ represents the Hubble parameter, $\rho_{\gamma}$ is the energy density of photon at temperature $T$, 
$J_i$ are the visible functions given in Ref.~\cite{McDermott:2019lch}.
The combined factor $\int d^3 p/(2\pi)^3 f_{p} P_{\rm osc} \Gamma_{\rm sca} m_{A'}$ represents the energy density injected into the photon plasma per unit time.
The lower and upper limits on the EM energy density injection can be derived from the parameters $y$ and $\mu$ \cite{Chluba:2011hw, Khatri:2012tw, Chluba:2013vsa, Tashiro:2014pga}, where 
the CMB distortion constraints requires $|y| < 1.5 \times 10^{-5}$ and $|\mu| < 6 \times 10^{-5}$ \cite{Tashiro:2014pga}. 
As a result, the CMB distortion provides a fairly strong constraints (labeled as CMB $\text{w}/\chi$), as shown by the dashed (solid) light brown line in Fig.~\ref{fig: result} for the two benchmark points, $m_\chi = 1$ keV  and $m_\chi = 1$ MeV. If there is no $\chi$ dark matter, the CMB limit is described by the gray shaded region, with label CMB.

Lastly, we consider other astrophysical constraints on the millicharged CoIDM. We choose the Bullet Cluster as a generic example following Ref.~\cite{Liu:2019bqw}. For CoIDM, the constraint is that the momentum exchange rate of $A'-\chi$ collisions should meet the condition that the dominant dark matter component has an effective hard collision rate  $\Gamma^{\rm{eff}}_{\rm{bullet}} \lesssim 0.016 \; \rm{Gyr}^{-1}$, where $\Gamma^{\rm{eff}}_{\rm{bullet}} = \Gamma_{\rm{sca}} \;m_{A'}/m_{\chi}$ in this model~\cite{Liu:2019bqw}. The additional mass suppression factor comes from the inverse scattering since we consider the momentum exchange rate. Because in the Bullet Cluster, the $A'$ and $\chi$ particles that collide with each other, originate from two separate DM halos. Their relative velocity $v_{\rm rel}^{\rm BC}\sim 4000 ~\rm{km/s}$ is much larger than their velocity dispersion $v_{0}^{\rm BC}\sim 1000 ~\rm{km/s}$ in the individual halo. Since the out-coming $A'$ will have a velocity of order $\mathcal{O}(v_{\rm rel})$ for a typical collision, there will be an exponential suppression in the Bose enhancement factor $\langle f_{A'} \rangle e^{-(v_{\rm rel}^{\rm BC}/v_{0}^{\rm BC})^2} \sim 10^{-7} \langle f_{A'} \rangle$. Furthermore, the DM density of the Bullet Cluster, $\rho^{\rm BC} \sim 10^{-2} \rm{GeV/cm^{-3}}$ is smaller than that of our galaxy. To evade the Bullet Cluster constraint, the interaction strength  in our model should satisfy $g_d \lesssim 0.1$ for MeV $\chi$, while for keV $\chi$ the constraint is strengthened to $g_d \lesssim 0.001$.

\section{conclusion}
\label{conclusion} 

We investigated the millicharged CoIDM model, which comprises two components dark matter: the dark photon $A'$ and the hidden Dirac fermion $\chi$. 
In this model, the dark photons and EM photons are coupled through mass mixing, and there is an unsuppressed vector interaction between $\chi$-$A'$, resulting a millicharged vector interaction between $\chi$-$A$.
Our findings reveal that even for a small fraction of the $\chi$ dark matter (with $f_\chi=10^{-7}$ and $f_\chi=10^{-10}$ for our two benchmarks), the scattering rate $\Gamma_{\rm sca}$ for the $A'-\chi$ scattering process exhibits significant Bose enhancement for light dark photons (i.e., $m_{A'} \sim 10^{-6}$ eV for radio telescopes). 
We observe that within the parameter space where $\Gamma_{\rm sca} \sim m_{A'}$, the oscillation conversion rate from DPDM to photon signals experiences a significant enhancement in the galaxy, thereby amplifying the photon signal.
In the benchmark presented in Fig.~\ref{fig: result}, the millicharged CoIDM model achieves sensitivities with radio telescopes such as LOFAR and SKA that can surpass the existing experimental limits by $6\sim 12$ orders of magnitude, with the presence of millicharged $\chi$ dark matter.
In addition, we also explore the constraints from the CMB spectral distortion on the millicharged CoIDM model.
Furthermore, we would like to note that the Bose enhancement in dark photons oscillating into photons bears similarity to the phenomenon of Pauli blocking in the scenario of active neutrino oscillations into sterile neutrinos in dense media.
Lastly, our discussion on the enhancement of dark photon oscillation to photons through increased scattering rates holds potential implications for the detection of other similar new physics phenomena.

\section{acknowledgement}

The work of JL is supported by the National Science Foundation of China under Grant No. 12075005, 12235001. 
The work of X.P.W. is supported by the National Science Foundation of China under Grant No. 12005009, No. 12375095 and the Fundamental Research Funds for the Central Universities.

\bibliography{ref.bib}{}
\bibliographystyle{utphys28mod}


\end{document}